\documentstyle[12pt, epsfig]{article}

\def\ba{\begin{eqnarray}}
\def\ea{\end{eqnarray}}
\def\be{\begin{equation}}
\def\ee{\end{equation}}
\def\l{\left}
\def\r{\right}
\def\dd{\partial}

\def\32{\frac{3}{2}}
\def\12{\frac{1}{2}}
\def\1N{\frac{1}{N}}

\def\vk{\vec k}

\def\vx{\vec x}
\def\vy{\vec y}
\def\d3{d^3\!}
\def\tin{t_{in}}

\newcommand{\cd}{\cdot}

\newcommand{\De}{\Delta}

\newcommand{\ra}{\rightarrow}
\newcommand{\bea}{\begin{eqnarray}}
\newcommand{\eea}{\end{eqnarray}}
\newcommand{\bean}{\begin{eqnarray*}}
\newcommand{\eean}{\end{eqnarray*}}

\newcommand{\rer}[1]{Eq.~(\ref{#1})}

\begin{document}

\begin{center}
{\Huge Cosmic Microwave Background Anisotropies Induced by Global
Scalar Fields: The Large $N$ Limit} \vspace{2cm}\\
{\large Martin Kunz and Ruth Durrer}\\
Universit\'e de Gen\`eve, Ecole de Physique, \\24, Quai E. Ansermet,
CH-1211 Gen\`eve, Switzerland
\end{center}
\begin{abstract}
We present an analysis of CMB anisotropies induced by global scalar
fields in the large $N$ limit. In this limit, the CMB 
anisotropy spectrum can be determined without cumbersome 3D
simulations. We determine the source functions and their unequal time
correlation functions and show that they are quite similar to the
corresponding functions in the texture model. This leads us to the
conclusion that the large $N$ limit provides a 'cheap approximation'
to the texture model of structure formation.
\end{abstract}
PACS numbers: 98.80.Cq 98.65.Dx
\vspace{0.3cm}

\noindent
The anisotropies in the cosmic microwave background (CMB) have become
an extremely valuable tool for cosmology. There are hopes that the
measurements of the CMB anisotropy spectrum might lead to a
determination of cosmological parameters like $\Omega_0,~ H_0,
~\Omega_B, ~\Lambda$ to within a few percent. The justification of this hope
lies to a big part in the simplicity of the theoretical
analysis. Fluctuations in the CMB can be determined almost fully
within linear cosmological perturbation theory and are not severely
influenced by nonlinear physics. 

Presently there are two competing classes of models which lead to a
Harrison Zel'dovich spectrum of fluctuations: Perturbations may be
induced during an inflationary epoch or they may be due to
scaling seeds like, e.g., a self ordering global scalar field or cosmic
strings (for a  general definition of 'scaling seeds' see \cite{draft}). 
In the first
class, the linear perturbation equations are homogeneous. In the
second class they are inhomogeneous, with a source term due to the
seed. The evolution of the seed is in general non--linear and
complicated and therefore much less accurate predictions have been
made so far for models where perturbations are induced by seeds.

In this communication we discuss an especially simple model
with seeds where the equation of motion for the seed perturbations can
be solved explicitly. We consider a $N-$component real scalar field
$\phi$ with  $O(N)$ symmetric potential $V$, which at $T=0$ is given
by $V=\lambda(\phi^2-\eta^2)^2$. At low temperatures, $T\ll\eta$,
$\phi$ can be regarded as constrained to an $N-1$ sphere with
radius $\eta$. The scalar field then evolves according to the
non--linear $\sigma$--model which is entirely scale
free. In terms of the dimension-less variable $\beta=\phi/\eta$ we find
\be
   	\Box\beta - (\beta\cd\Box\beta)\beta =0 ~,\label{si}
\ee
with the condition $\beta^2=1$.
The non--linearity in this equation, 
$-(\beta\cd\Box\beta)\beta= (\dd_\mu\beta\cd\dd^\mu\beta)\beta$
contains a sum over $N$ components. In the limit $N\rightarrow
\infty$, this sum can be replaced by an ensemble average and the
resulting linear equation of motion can be solved exactly. One obtains
 \cite{TS}
\be
	\beta(\vec k,t)=At^{\frac{3}{2}}
	\frac{J_\nu (kt)}{(kt)^\nu}\beta_{in} (\vec k).  \label{beta}
\ee
The index $\nu$ is determined by the background matter model and
varies between $\nu= 2$ in a radiation dominated background and $\nu =
3$ in a matter dominated background. The pre-factor $A$ is chosen to
ensure $\beta^2=1$, 
\[ A = \left\{\begin{array}{ll}
	15/16 ,& \mbox{if } \nu=2,\\
	2835/128 ,& \mbox{if } \nu=3.\end{array} \right.	\]
The components of $\beta_{in}$ are assumed to be independent,
Gaussian-distributed random variables with vanishing mean and
dispersion $\langle(\beta_{in})_j^2\rangle=1/N$
for all values of $j$. (Clearly, the variables
$(\beta_{in})_j$ cannot be completely independent, since they obey the
condition $\sum_j(\beta_{in})_j^2 =1$.)

Once the scalar field $\beta$ is known, we can calculate its
energy--momentum tensor, the induced gravitational field and and its
action on matter and radiation within linear cosmological perturbation
theory. As has been discussed in \cite{TS}, the energy density of
a four component global scalar field is already quite close to the
large $N$ limit and there are thus justified hopes, that this simple
model might provide a quite sensible approximation to the texture
scenario for structure formation. On the other hand, we know that
non--linearities, which lead to the mixing of scales and to the
deviations from a Gaussian distribution,  are crucial for some 
qualitative properties of defect models, like decoherence 
\cite{joao,draft}. In the large$-N$ limit,
the only non--linearities are the quadratic expressions of the energy
momentum tensor, and thus effects like decoherence might be mildened
substantially in this model.

This communication is an outline of a longer paper in preparation. We
first briefly repeat the basic equations for the determination
of the CMB anisotropy spectrum in the presence of seeds, and  solve
the equations (for given seed functions) in a simplified situation.
We then outline the calculation of the relevant correlation functions and
present some results. We end with conclusions and an outlook.
\vspace{0.5cm}

The coefficients of the angular power spectrum
of CMB anisotropies are related to the 2--point function according  
to~\cite{Paddi} 
\be
\l. \l<\frac{\delta T}{T}(\vec n)\frac{\delta T}{T}(\vec n')\r>
\r|_{(\vec n \cdot \vec n'=\cos\varphi )}
=\frac{1}{4\pi}\sum_\ell (2\ell+1)C_\ell P_\ell(\cos \theta) ~.
\ee
For pure scalar perturbations, neglecting Silk damping, the $C_\ell$'s
are given by\cite{draft}

\be
C_\ell=\frac{2}{\pi}\int\frac{\l<\l|\Delta_\ell(\vk)\r|^2\r>}{(2\ell+1)^2}
k^2dk    \label{Cl}
\ee
where
\be
\frac{\Delta_\ell}{2\ell+1}=\frac{1}{4}\Delta_{gr}
(\vk,t_{\mbox{\scriptsize dec}})j_\ell(kt_0)-
V_r(\vk,t_{\mbox{\scriptsize dec}})j_\ell '(kt_0)
+k \int_{t_{d\!e\!c}}^{t_0} (\Psi-\Phi)(\vk,t') j_\ell '(k(t_0-t'))dt'.
	\label{Dl}\ee

For large $\ell$ this spectrum is corrected by Silk damping, which can
be approximated by multiplying $\De_\ell(k)$ with an exponential
damping envelope \cite{HuSu}.

We want to consider the situation where fluctuations are induced by
seeds. We restrict ourselves to scalar perturbations. The energy
momentum tensor of scalar seed perturbations can be parameterized by
the following 
four functions: $f_\rho$ the energy density of the seed, $f_p$ the
pressure of the seed, $f_v$ a potential for the energy flux of the
seed and $f_\pi$ the potential for anisotropic stresses of the
seed~\cite{d90,ruth} (see below). The linear cosmological 
perturbation equations are then of the form
\be
	{\cal D}X_j=M_j^iF_i ~,  \label{Deq}
\ee
where ${\cal D}$ is a first order linear differential operator, $X$ 
is a vector consisting of all, say $m$, gauge invariant perturbation 
variables of the cosmic
fluids (like $\De_{gr}$, $V_r$, and, e.g. the corresponding variables 
for the cold dark matter (CDM), ...). $F=(f_\rho,f_p,f_v,f_\pi)$ is the 
source vector and $M$ is a, in general time dependent, $m\times 4$ matrix.

The general solution to \rer{Deq} is of the form
\be X_j(t)=\int_{t_{in}}^tG_j^i(t,t')F_i(t')dt' ~,\ee
where $G$ is the Green's function of the differential operator $\cal D$.
The Bardeen potentials $\Phi$ and $\Psi$ are algebraic
combinations of the fluid variables $X_j$ and the source functions
$F_i$, 
\be  (\Psi-\Phi)(t) = P^j(t)X_j(t)+Q^i(t)F_i(t) ~.\ee

Inserting this solution in \rer{Dl} we obtain
\bea
{\Delta_\ell\over 2\ell+1} &=& \int_{\tin}^{t_{dec}}dt\left[{1\over
4}G_{\De_r}^i(t_{dec},t)j_\ell(kt_0) -G_{V_r}^i
(t_{dec},t)j'_\ell(kt_0)\right]F_i(t) \nonumber \\
&&
   + k\int_{t_{dec}}^{t_0}dt\left[\int_{\tin}^tdt'(P^j(t)G_j^i(t,t')F_i(t'))
    ~ + ~  Q^i(t)F_i(t)\right] ~.
\eea
The expectation value in \rer{Cl} thus consists of time integrals of 
unequal time correlations  of the source functions, which are of the
generic form
\bea
{\langle|\Delta_\ell|^2\rangle\over (2\ell+1)^2} &=&
 \int^{t_{dec}}_{\tin}dt\int^{t_{dec}}_{\tin}dt'
	\langle F_i(t)F_j(t')\rangle A_\ell^{ij}(t,t') ~+~
\nonumber \\  &&
~ + ~  \int_{t_{dec}}^{t_0}dt \int_{t_{dec}}^{t_0}dt'
        \int_{t_{in}}^{t}dt''\int_{t_{in}}^{t'}dt'''
	\langle F_i(t'')F_j(t''')\rangle B_\ell^{ij}(t,t',t'',t''') ~ + ~ 
\nonumber \\  &&
~ + ~   \int_{t_{dec}}^{t_0}dt'\int_{t_{dec}}^{t_0}dt 
        \int_{t_{in}}^{t}dt''
	\langle F_i(t')F_j(t'')\rangle C_\ell^{ij}(t,t',t'') ~+~
\nonumber \\  &&
~ + ~   \int^{t_{dec}}_{\tin}dt' \int_{t_{dec}}^{t_0}dt
        \int_{t_{in}}^{t}dt''
	\langle F_i(t')F_j(t'')\rangle D_\ell^{ij}(t,t',t'') ~ + ~ 
\nonumber\\  &&
~ +~  \int_{t_{dec}}^{t_0}dt\int_{t_{dec}}^{t_0}dt'
	\langle F_i(t)F_j(t')\rangle E_\ell^{ij}(t,t') ~ + ~
\nonumber\\  &&
~ +~  \int_{t_{dec}}^{t_0}dt\int^{t_{dec}}_{\tin}dt'
	\langle F_i(t)F_j(t')\rangle H_\ell^{ij}(t,t')  ~.
\eea
To calculate the $C_\ell$'s we need therefore need to know the unequal
time correlations of the seed functions $F_i$ and the Green's
functions for the cosmological model specified. In general this is a
quite formidable task. Here we shall just discuss  a toy
model. A somewhat more complicated example is given in~\cite{joao}.

We consider a pure radiation universe with vanishing spatial curvature.
In this case, the linear perturbation equations are given by~\cite{ruth}
\ba
\Phi &=& \frac{1}{x^2+6}\l(x^2 \Phi_S+\32 \Delta_{gr}+
	6 \frac{V_r}{x} \r) \label{phi}\\
\Psi &=& -\Phi-2 \epsilon f_\pi  \label{psi}\\
\Delta_{gr}' &=& -\frac{4}{3} V_r\\
V_r' &=& \Psi-\Phi+\frac{1}{4} \Delta_{gr}
\ea
where $x=kt$ and a prime denotes a derivative w.r.t $x$. The energy 
momentum tensor of the source enters via the combinations
$f_\pi$ and $\Phi_S=\epsilon(f_\rho/k^2+3 f_v/(kx))$.
These equations can be combined to a second order differential-equation for
$\Delta_{gr}$ alone,
\be
\Delta_{gr}''+\frac{12}{x^2+6} \frac{\Delta_{gr}'}{x}
+\frac{1}{3} \frac{x^2-6}{x^2+6} \Delta_{gr} =
\frac{8}{3}\l(\epsilon f_\pi + \frac{x^2}{x^2+6} \Phi_S\r) \label{DGL}~,
\ee
with homogeneous solutions
\[
D_1(x) =  \cos\l(\frac{x}{\sqrt{3}}\r)
-2 \frac{\sqrt{3}}{x}\sin\l(\frac{x}{\sqrt{3}}\r) ~,~~
D_2(x) =  -\sin\l(\frac{x}{\sqrt{3}}\r)
-2 \frac{\sqrt{3}}{x}\cos\l(\frac{x}{\sqrt{3}}\r)~,
\]
leading to the Greens function
\be
G(x,x')={\sqrt{3}x'\over x(6+x'^2)}\left[(12+xx')\sin({x-x'\over\sqrt{3}})
	+2\sqrt{3}(-x+x')\cos({x-x'\over\sqrt{3}})\right] ~.
\ee
On super-horizon scales ($x\ll 1$) the solutions of the homogeneous
equations consist of one
 constant and one decaying, $\propto 1/x$, mode,  while for $x\gg 1$
we obtain two oscillating modes. The general solution
with source term $S(x)$ and initial condition $\De_{gr}(0)=V_r(0)=0$ is 
\bea
\Delta _{gr}(x) &=& \int_0^xG^1(x,x')S(x')dx'\\
V_r(x)          &=& \int_0^xG^2(x,x')S(x')dx'   ~~~\mbox{ where}\\
 G^1 = G & \mbox{ and }& G^2=-{3\over 4}{dG\over dx} ~.
\eea
Together with \rer{phi} and (\ref{psi}) it is now straight forward
 to determine
the integral kernels A,B,C,D,E and H and thus, for given source
correlation functions, the $C_\ell$'s.

Let us therefore discuss the source
correlation functions of scalar field sources in the large $N$ limit.
The seed functions  are given by~\cite{ruth}:

\ba
f_\rho = F_1 &=& \frac{1}{2} \l[\dot{\beta}^2+(\nabla\beta)^2\r]\\
f_p = F_2 &=& \frac{1}{2} \l[\dot{\beta}^2 -\frac{1}{3} (\nabla\beta)^2\r]\\
f_v =F_3&=& \Delta^{-1} \l[\dot{\beta}\cdot \beta_{,j}\r]^{,j}\\
f_\pi =F_4&=& \frac{3}{2} \Delta^{-2} \l[\beta_{,i}\cdot\beta_{,j}
	-\frac{1}{3}\delta_{ij}(\nabla\beta)^2\r]^{,ij}
\ea

Using the fact that the initial fields are  uncorrelated and Gaussian
distributed, 
\be
\l<\beta_i(\vec k) \beta_j (\vec p)\r> = \frac{1}{N} \delta_{ij}
\delta(\vec k+\vec p)
\ee
and using the exact solution \rer{beta}
we find the power spectra and the unequal time correlation functions 
of the seed variables $F_i$. Below we give explicit expressions for
the power spectra of $f_\rho~,f_v$ and $f_\pi$ and, as an example, the unequal
time correlation function for $f_v$. These integrals can be evaluated
numerically, examples of which are shown in Figs.~1 and 2.
We define $\vx=t\vk$ and we  set
\be
\chi(x) \equiv \frac{J_\nu(x)}{x^\nu} ~,~~~ 
\varphi(x) \equiv {3\over 2}\chi(x)-\frac{J_{\nu+1}(x)}{x^{\nu-1}}~.
\ee
Using these abbreviations we obtain the somewhat cumbersome expressions
\ba
\lefteqn{\l<f_\rho(\vk,t)\cdot f_\rho(\vk ',t)\r> =} \nonumber \\
&& \frac{A^2}{t}\frac{\delta(\vk+\vk ')}{2N} \int \d3 y 
\l\{ \varphi(y)^2  \varphi(|\vy-\vx|)^2 +
\l[\vy (\vx-\vy)\r]^2 \chi(y)^2 \chi(|\vy-\vx|)^2 \r. ~, \nonumber \\
&&\l. -2 \vy(\vx-\vy) \varphi(y)\varphi(|\vy-\vx|)
	\chi(y) \chi(|\vy-\vx|) \r\} \label{frho}\\
\lefteqn{\l<f_v(\vk,t)\cdot f_v(\vk ',t)\r> =} \nonumber \\
&&A^2t\frac{\delta(\vk+\vk ')}{N}\int \d3 y \frac{\vx(\vx-\vy)}{x^4}
	\varphi(y)\chi(|\vy-\vx|) \nonumber \\
&&\l[\vx(\vx-\vy)\varphi(y)
\chi(|\vx-\vy|)+\vx\vy\varphi(|\vx-\vy|)\chi(y)\r]~,\\
\lefteqn{\l<f_\pi(\vk,t)\cdot f_\pi(\vk ',t)\r> =} \nonumber \\
&& A^2t^3 \frac{9\delta(\vk+\vk ')}{2N} \int \d3 y
\frac{\l[ (\vx\vy)(x^2-\vx\vy)+\frac{1}{3} x^2 (y^2-\vx\vy)\r]^2}{x^8}
\chi(y)^2 \chi(|\vx-\vy|)^2  \label{fpi} ~,\\
\lefteqn{\l<f_v(\vk,t)\cdot f_v(-\vk ',t')\r>= }\nonumber\\
&&\frac{A^2 t}{N} \frac{r^2}{x^4} \delta(\vk - \vk ') \int d^3y \l\{
\l[x^2-\vec x \vec y\r]^2\varphi(y)\chi(|\vec x-\vec y|)
\varphi(yr)\chi(|\vec x-\vec y|r)\r.\nonumber \\
&&+\l.\l[(x^2-\vec x \vec y)(\vec x \vec y)\r]
\varphi(y)\chi(|\vec x-\vec y|)
\varphi(|\vec x-\vec y|r)\chi(yr)\r\} ~,
\ea
where we have set $r=t'/t$ in the last equation.
The behavior of these functions on very large and very small scales
can be obtained analytically. On super horizon scales, $x\ra 0$, the power
spectra for $f_\rho~,~f_p$ and $f_v$ behave like white
noise. Numerically we have found
\bea
\l<|f_\rho|^2\r> &{\ra}_{x\ra 0}& {1\over Nt} \cdot \l\{ \begin{array}{rc}
 9.36  \cdot 10^{-2} & \nu = 2\\
       10.31 \cdot 10^{-2} & \nu = 3 \end{array}\r. \\
\l<|f_p|^2\r> &{\ra}_{x\ra 0}& {1\over Nt}  \cdot \l\{ \begin{array}{rc}
   1.066 \cdot 10^{-2} & \nu = 2\\
      0.805 \cdot 10^{-2}  & \nu = 3 \end{array}\r. \\
\l<|f_v|^2\r> &{\ra}_{x\ra 0}& {t\over N} \cdot \l\{ \begin{array}{rc}
	  0.8687 \cdot 10^{-2} & \nu = 2\\
      0.4694 \cdot 10^{-2} & \nu = 3 \end{array}\r. \\
\eea

From general arguments \cite{joao,draft}, we would have expected
also $f_\pi$ to behave like white noise on super--horizon scales. 
However, from \rer{fpi} we find that $f_\pi$ diverges at small $x$ like
$1/x^2$. Even though we do not quite understand this result, it does
not lead to divergent Bardeen potentials, if we allow for anisotropic
stresses in the  matter (like e.g. from a component of massless
neutrinos). In this case it can be shown \cite{draft} that
compensation arranges the anisotropic stresses in the fluid, $p\Pi$,
such that $f_\pi+p\Pi \propto x^2f_\pi$. Therefore, also the
anisotropic stresses contribute a white noise component to
  the Bardeen potentials on super-horizon scales, namely:
 \be
 x^4 \l<|f_\pi|^2\r> {\ra}_{x\ra 0}  
	{ t^3 \over N} \cdot \l\{ \begin{array}{rc}
       5.169 \cdot 10^{-2} & \nu = 2\\
       6.539 \cdot 10^{-2} & \nu = 3 \end{array}\r.
 \ee

In the limit $x\ra\infty$ the source functions decay like  
\bea
\l<|f_\rho|^2\r>~,~~ \l<|f_p|^2\r>&{\ra}_{x\ra \infty}& 
	{ x^{1-2\nu} \over Nt}  \\
\l<|f_v|^2\r> &{\ra}_{x\ra \infty}& { x^{-1-2\nu} t\over N}  \\
\l<|f_\pi|^2\r> &{\ra}_{x\ra \infty}& { x^{-3-2\nu} t^3\over N}  \\
\eea
In Fig.~1 we plot $t<|f_\rho|^2>(x)$ as obtained from 
\rer{frho}, and compare it to the corresponding function found by 3D 
simulations of the texture model.

The normalized unequal time correlation functions are defined by
\be
C_i(k,t,t')={\langle f_i(k,t)f^*_i(k,t')\rangle\over 
      \sqrt{\langle|f_i(k,t)|^2\rangle\langle|f_i(k,t')|^2\rangle}}
\ee
In the large $N$ limit, the correlation functions decay like power
laws. For $r\equiv t'/t$ we find in the limit $r\gg 1~,~ kt'\gg 1$
the behavior $C_i \propto r^{-\gamma_i}$, with
\be
 \gamma_\rho = 3/2 ~,~~
 \gamma_p = 3/2 ~,~~
\gamma_v = 3/2 ~,~~
\gamma_\pi = 5/2  ~.
\ee
It is not quite clear to us, whether this behavior is reproduced in
the texture model. Due to the arguments given at the
beginning, it may well be that $N=4$ and $N\ra\infty$ show a different
decoherence behavior. Originally (taking into account the numerical
accuracy of about 10\% of the 3D simulations) we approximated the
decoherence in the texture model with an  exponential decay law. 
However, comparing the unequal time
correlation functions for $\dot{\beta}^2$ shown  in Figs.~2 and 3 
for the large $N$ limit and a 3D simulation of the texture model 
respectively, we realize, that they agree extremely well in the
numerically most reliable, central region, and the seemingly
stochastic higher order oscillations also found in the texture model
might actually be real (see Fig.~4), leading to power law decoherence.

Using these source functions, we have determined the CMB anisotropy
spectrum induced by the large $N$ limit of a self ordering scalar field
for a spatially flat cosmological model with CDM, radiation and
baryons. Since decoherence is so weak for large $N$, we used the
approximation of perfect coherence, $C_i\equiv 1$. This simplification
has been used so far for all 'analytic approximations' of $O(N)$
models and, e.g., in the case of textures, $N=4$, it seems to agree
reasonably with numerical simulations \cite{neil}.  This will
certainly be even more so in the large $N$ limit. The influence of
decoherence on the CMB power spectrum is discussed in \cite{joao}.
\vspace{1cm}

We have shown that the large $N$ limit of global scalar fields
provides a model of seeded structure formation where CMB
anisotropies can be determined without cumbersome numerical
simulations and thus with much higher accuracy and large dynamical
range at relatively modest costs. Determining the correlation
functions of the seed variables just requires numerical
convolution of Bessel functions multiplied with powers. Once the seed
correlation functions are known, perturbations in matter and radiation
can be calculated by solving a system of linear perturbation
equations, very similar to the homogeneous case of inflationary
perturbations.

We believe that the large $N$ limit has many features in common with
the texture model of structure formation and thus provides a ``cheap
approximation'' to this model. The most obvious difference between
the analytic limit and the texture model is the decoherence
behavior. In the large $N$ limit, the field evolution is linear and
non--linearities, which are responsible for decoherence, enter only via 
the quadratic expressions of the energy momentum tensor. We thus
expect decoherence to be somewhat weaker in the large $N$ limit.

In a forthcoming paper, we plan to work out the large $N$ limit in
more detail, and to study the dependence of the resulting CMB anisotropy
spectrum of cosmological parameters. We also want to investigate more
fully the comparison of the large $N$ source functions with the source
functions found in 3D simulations of the texture model.
The limit discussed here provides an very
useful toy model for structure formation with scaling seeds for which
decoherence is not important.

\newpage
\begin{figure}[!htb]
\epsfig{file=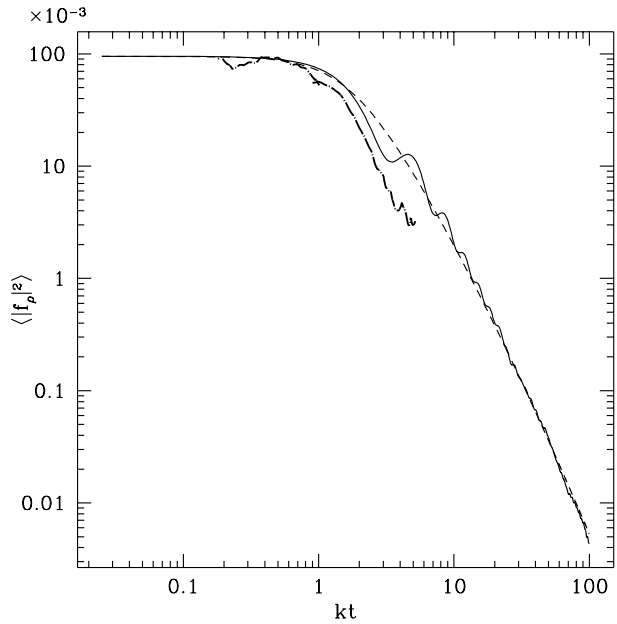, width=8.5cm}
\caption[Figure1]{The source function $f_\rho$: In the large $N$ limit
        (full line), the approximation used to calculate the $C_\ell$'s
        in Fig. 5 (dashed line) and from a 3 dimensional numerical 
        computation (dot-dashed line) for the texture model ($N=4$).}
\label{fig1}
\end{figure}

\begin{figure}[!htb]
\epsfig{file=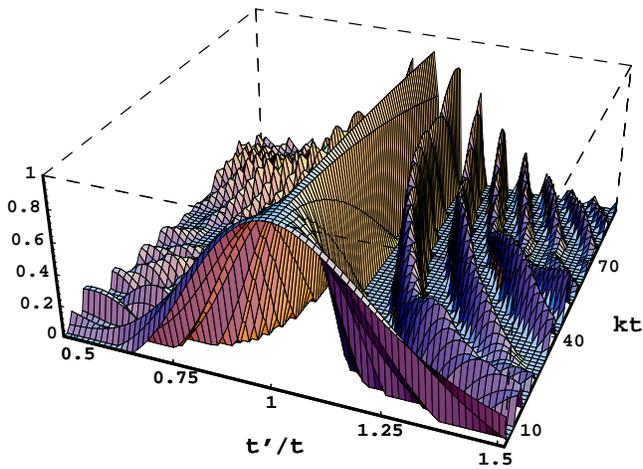, width=8.5cm}
\caption[test]{The unequal time correlation function for
$\dot{\beta}^2=f_\rho+3f_p$ at fixed $t$ as function of $t'$ and
$k$ in the large $N$ limit. Negative values are set to zero.}
\label{fig2}
\end{figure}

\begin{figure}[!htb]
\epsfig{file=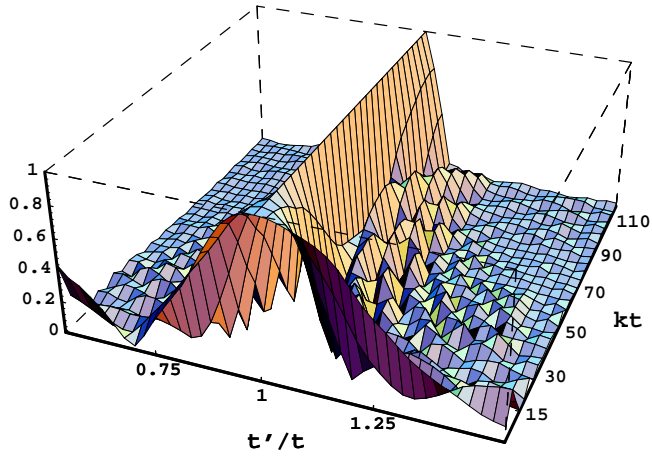, width=8.5cm}
\caption[test]{The same as Fig.~2 for the texture model. The
similarity is obvious. The high order oscillations which are very
pronounced in the large $N$ limit are washed out or absent in the
texture model. Whether this is a real feature or just numerical
inaccuracy or both is not yet clear.}
\label{fig3}
\end{figure}

\begin{figure}[!htb]
\epsfig{file=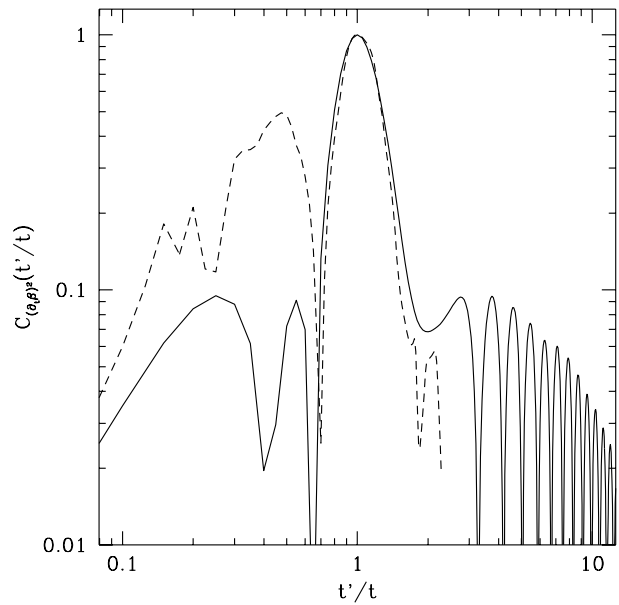, width=8.5cm}
\caption[test]{A cut through Fig.~2 (solid line) and Fig.~3 (dashed
line) at $kt=3.9$. The central peaks are in very good
agreement. Secondary peaks do not agree and the decay law for the
texture model is difficult to predict from this data.
}
\label{fig4}
\end{figure}

\begin{figure}[!htb]
\epsfig{file=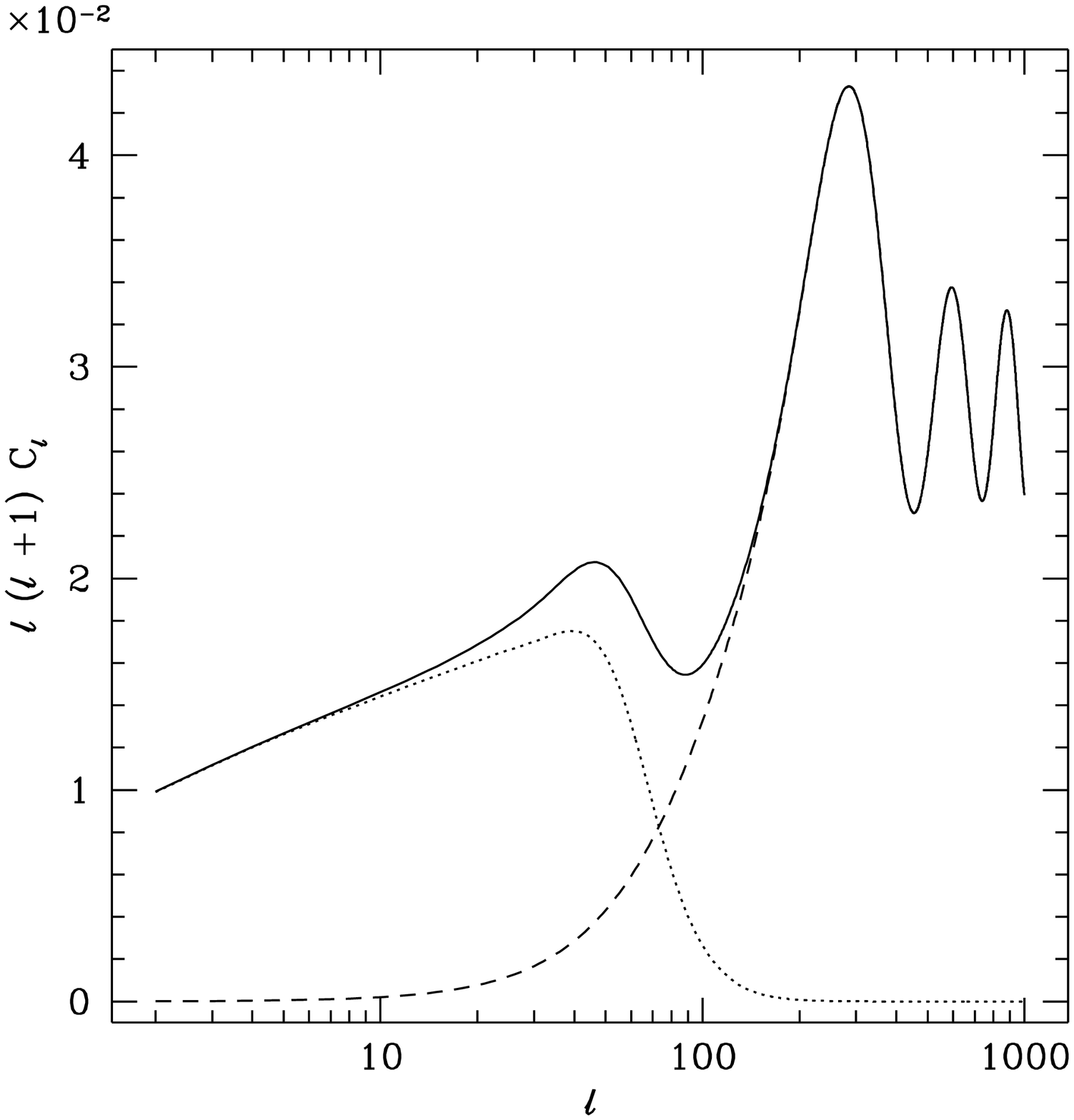, width=8.5cm}
\caption[Figure 5]{The CMB anisotropy spectrum, obtained  by using
          polynomial fits
        for the source functions as shown in Fig. 1. Only scalar
          perturbations are included. The Sachs Wolfe part is
          indicated by the dotted line, the dashed line represents
	the acoustic contributions. Silk damping is not included.}
\label{fig5}
\end{figure}


\begin{thebibliography}{99}
\bibitem{draft}R. Durrer, M. Sakellariadou and M. Kunz, in preparation (1996).
\bibitem{TS}N. Turok and D. Spergel,
	{\em Phys. Rev. Lett.} {\bf 66}, 3093 (1991).
\bibitem{joao}J. Magueijo, A. Albrecht, P. Ferreira and D. Coulson,
	preprint, archived under astro-ph/9605047 (1996).
\bibitem{HuSu}W. Hu and N. Sugiyama, ``Small scale cosmological perturbations:
	an analytic approach'', astro-ph/9510117 (1996).
\bibitem{Paddi}T. Padmanabhan, {\em Structure Formation in the
	Universe}, Cambridge University Press (1993).	
\bibitem{d90}R. Durrer, {\em Phys. Rev.} {\bf D42}, 2533 (1990).
\bibitem{ruth}R. Durrer,
        {\em Fund. of Cosmic Physics} {\bf 15}, 209, (1994).
\bibitem{neil}N. Turok, Preprint, archived under astro-ph/9600687.




\end{thebibliography}
\end{document}